\documentclass[twocolumn,english,aps,showpacs,superscriptaddress]{revtex4-1}
\usepackage{amsmath,amssymb}
\usepackage{graphicx}
\usepackage{txfonts}
\usepackage{color}
\usepackage{hyperref}
\usepackage{bbold}
\usepackage{framed}
\usepackage{wrapfig}
\usepackage[letterpaper]{geometry}
\begin{document}

\title{Specific heat in strongly hole-doped Iron-based  superconductors}

\author{Dmitry V. Chichinadze and Andrey V. Chubukov}
\affiliation{School of Physics and Astronomy,
University of Minnesota, Minneapolis, MN 55455, USA}

\begin{abstract}
We compute specific heat $C(T)$ in a
  strongly hole-doped Fe-based superconductor, like KFe$_2$As$_2$, which has
  only hole pockets. We model the electronic structure by
   a three-orbital/three pocket model with two smaller hole pockets  made out of $d_{xz}$ and $d_{yz}$ orbitals and a larger pocket made out of $d_{xy}$ orbital.  We use as an input the experimental fact that the mass of $d_{xy}$  fermion is
      several times  heavier than that of $d_{xz}/d_{yz}$ fermions.   We argue that the heavy $d_{xy}$ band
        gives the largest contribution  to the specific heat in the  normal state, but the superconducting gap on the $d_{xy}$ pocket is  much
           smaller than that on $d_{xz}/d_{yz}$ pockets. We argue that in this situation the jump of $C(T)$ at $T_c$ is  determined by $d_{xz}/d_{yz}$ fermions, and the ratio $(C_s-C_n)/C_n$ is a fraction of that  in a one-band BCS  superconductor. At  $T<T_c$,  $C(T)$ remains relatively flat  down to some $T^*$, below which it  rapidly drops.
   This behavior is consistent with the data for KFe$_2$As$_2$ and related materials.  We use one-parameter model for the interactions and fix
    this only parameter by matching the experimental ratio of the gaps on the two $d_{xz}/d_{yz}$ pockets. We argue that the resulting parameter-free model reproduces quantitatively the data on $C(T)$ for KFe$_2$As$_2$.    We further argue that the very existence of a finite $T^* < T_c$  favors $s^{+-}$ gap structure over $d-$wave, because in the latter case $T^*$ would almost vanish.
\end{abstract}

\maketitle

{\it \bf Introduction.}~~~
Rich physics of Iron-based superconductors (FeSC) continues  to attract strong attention from the condensed-matter community
\cite{Ishida2009,Johnston2010,mazin2010,paglione2010high,Wen2011,basov2011,Hirschfeld2011,Stewart2011,Chubukov2012Review,Bascones_review,Chubukov_Hirschfeld,Graser2009,fernandes2014drives}.
One of the most debated issues in the field is the strength of correlations. On one hand,  FeSCs have Fermi surfaces,  and most display a metallic, Fermi-liquid like  behavior in some temperature range above superconducting $T_c$.  On the other, there is a clear distinction between the observed
  electronic structure and the one obtained by first-principle calculations for free fermions.
  Some researchers believe that this difference can be accounted for by including the momentum-dependent
  self-energy~\cite{Benfatto},
   which modifies the dispersion but leaves fermions and their collective degrees of freedom fully coherent
   (this is often termed as "itinerant scenario", see e.g.,
    Ref. \cite{Chubukov2012Review,Chubukov_book}). Others argue that at  energies relevant to superconductivity and competing orders, fermions can be viewed as correlated yet itinerant, but collective magnetic excitations should be viewed as at least partly localized    (a "Hund metal scenario",  see, e.g., Ref. \cite{Kotliar_11,pavarini2017}). And others further  argue  \cite{deMedici2014,Medici_book}   that electronic excitations should be viewed as  itinerant on some Fe-orbitals and as nearly localized on other orbitals (an "orbital selective Mottness" scenario).

  From the perspective of Mott physics, the best candidates to display Mott behavior are strongly hole-doped FeSCs, like KFe$_2$As$_2$ \cite{okazaki2012,Sato2009,Zabolotnyy2009,Terashima2010,Ota2014,Hardy2014}, as for these systems the tendency towards electron localization has been argued to develop at a smaller Hubbard $U$   (Ref. \cite{deMedici2014,Medici_book}).  Low-energy fermionic states in
 KFe$_2$As$_2$ are composed of fermions from three orbitals, $d_{xy}$, $d_{xz}$, and $d_{yz}$, the last two are related by $C_4$ symmetry \cite{okazaki2012}.
   Specific heat measurements in  KFe$_2$As$_2$ have shown that  above superconducting $T_c$, specific heat coefficient $C(T)/T$ scales as $a + bT^2$, as expected in a metal, but $a$ is larger than in other FeSCs \cite{Hardy2013, Hardy2014,Budko2012,Abdel-Hafiez2012,Fukazawa2009,Eilers2016}. Because $a$ is proportional to the sum of the effective masses for different bands, large value of $a$ implies that at least one  effective mass is large.
   Within the Mott scenario, the mass enhancement comes from frequency-dependent self-energy, $\Sigma (\omega)$. This self-energy narrows the dispersion and simultaneously reduces the quasiparticle residue $Z$, transferring $1-Z$ spectral weight into Hubbard sub-bands.  The effect is believed to be the strongest for the band made of fermions from $d_{xy}$  orbital     \cite{deMedici2014,Medici_book}.
       However,  band narrowing and accompanying mass enhancement  can be also caused by innocuous reasons like  smaller hopping integral for $d_{xy}$  fermions or closeness to a Van-Hove singularity (see~\cite{vh} and references therein). In the latter case, large value of the specific heat coefficient can be understood already within the itinerant scenario.
     ARPES data do indeed show~
     \cite{rhodes2018scaling,vh,Borisenko}
      that the $d_{xy}$ band is more narrow than the bands made by fermions from $d_{xz}$ and $d_{yz}$ orbitals,
      but Hubbard sub-bands have not been yet detected in  KFe$_2$As$_2$. Furthermore, some ARPES data
       on KFe$_2$As$_2$ and other FeSCs
       show that $d_{xy}$ excitations are as sharp as excitations
     from $d_{xz}/d_{yz}$ bands \cite{Sato2009,Ye2014}.
      This makes the interpretation of specific heat data above $T_c$ somewhat ambiguous.

\begin{widetext}

\begin{figure}[t]
\centering{}\includegraphics[width=0.99\columnwidth]{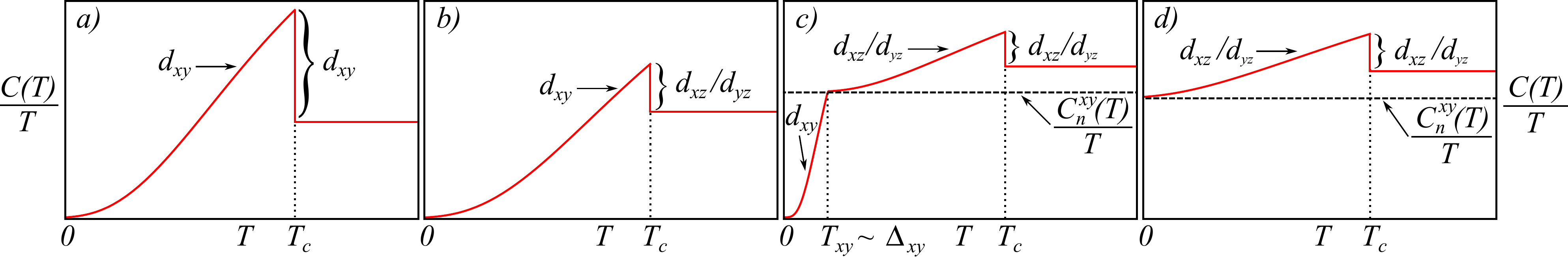}
\caption{Four different scenarios for the behavior of $C(T)/T$ in  the three-pocket model with light $d_{xz}/d_{yz}$ bands and a heavy $d_{xy}$ band.
(a) the specific heat both above and below $T_c$ is determined by the $d_{xy}$ band;
 (b) The specific heat jump is defined by the gap opening on $d_{xz}/d_{yz}$ bands, but $T$ dependence of $C(T)$ below $T_c$ is still determined by the $d_{xy}$ band; (c) The specific heat jump at $T_c$ and the behavior at $T_{xy}< T < T_c$ is determined by $d_{xz}/d_{yz}$ bands, while the contribution to $C(T)$ from the $d_{xy}$ band remains the same as in the normal state (the dashed line).  Below $T_{xy}$, the gap on the $d_{xy}$ band becomes larger than  $T$, and $C(T)/T$ rapidly drops; (d) the case when $T_{xy} =0$.}
\label{cplot}
\end{figure}

\end{widetext}

   In this communication we analyze whether one can separate between  Mott  and itinerant scenarios by analyzing specific heat data in the superconducting state. Given that $d_{xy}$ fermions have the largest mass, i.e., the largest density of states (DOS), there are four possibilities for system behavior below $T_c$. They are depicted in Fig. \ref{cplot}. One possibility (panel (a)) is that superconductivity predominantly develops on the   heavy  $d_{xy}$ orbital because of larger DOS. If this is the case, the system's behavior is the same as in a one-band superconductor:  the specific heat jump at $T_c$, $\delta C/C_n =(C_s-C_n)/C_n$, is of order one, and $C(T)$ varies as a function of a single variable $T/T_c$ below $T_c$. Another  (panel (b)) is that superconductivity develops at $T_c$ on $d_{xz}/d_{yz}$ orbitals, but the temperature dependence of $C(T)$  below $T_c$  is still determined by the heavy $d_{xy}$ orbital.  In this situation
     $\delta C/C_n$ is small, but $C(T)$ below $T_c$ is  the same as in panel (a). The third possibility  (panel (c))is that
        not only $(C_s-C_n)/C_n$ at $T_c$ but also the behavior of $C(T)$ in some $T$ range below $T_c$ is determined by $d_{xz}/d_{yz}$ orbitals, while  fermions on the  $d_{xy}$ orbital have smaller gap  and
          can be treated as non-superconducting down to $T_{xy} < T_c$. In this situation  $(C_s-C_n)/C_n$ is small, $C(T)/T$ varies slowly between $T_c$  and $T_{xy}$  towards a finite value (equal to normal state $C(T)/T$  for $d_{xy}$ fermions), and rapidly drops below $T_{xy}$. And the fourth possibility  (panel (d)) is that fermions on the $d_{xy}$ orbital do not pair down to $T=0$, i.e.,
     $T_{xy} =0$.

 The data for KFe$_2$As$_2$ from several groups
 ~\cite{Hardy2013, Hardy2014,Budko2012,Abdel-Hafiez2012,Fukazawa2009,Eilers2016} show that (i)  the specific heat jump at $T_c$  is much smaller than the BCS value, (ii) between $T_c$ and
    approximately $T_c/6$, $C(T)/T$ decreases rather slowly towards a finite value, (iii) below $T_c/6$, $C(T)/T$ rapidly drops and tends to zero at $T \to 0$. This behavior is consistent with the one in Fig.  \ref{cplot}(c).   We analyze whether this behavior can be understood by just assuming that the $d_{xy}$ band is heavier  than the other two bands (and, hence, the DOS for this band is the largest), or one needs to additionally include the reduction of quasiparticle $Z$ for the $d_{xy}$ band.   A momentum/frequency independent $Z$ can be absorbed into the renormalization of the interactions involving $d_{xy}$ fermions, hence the issue is whether mass/DOS variation between $d_{xy}$ and $d_{xz}/d_{yz}$ bands is sufficient to describe the data, or one needs to additionally assume that the interactions involving $d_{xy}$ fermions are weaker than the ones between $d_{xz}$ and $d_{yz}$ fermions.

We  argue that the difference in the masses is sufficient to describe the observed behavior.
 Namely, we obtain the behavior in Fig. \ref{cplot}(c)
 by analyzing the model of three $\Gamma-$ centered $d_{xz}/d_{yz}$ and $d_{xy}$ hole pockets in 2 Fe zone, and invoking mass difference but  keeping the interactions on all three orbitals comparable in strength.  If $Z$ on the $d_{xy}$ orbital is small in KFe$_2$As$_2$, this will  additionally  reduce  the value of $T_{xy}$.
  We note in passing  that our
  theoretical scenario is different from the one presented in Ref. \cite{Eilers2016}  as we do not require that KFe$_2$As$_2$ is close to a magnetic  quantum criticality.
   It is also different from the one in Ref. \cite{Hardy2013} where the temperature evolution of $C(T)$ was largely attributed to the gaps on hole barrels near $(\pi,\pi)$ in 2 Fe zone.  We emphasize that the existing ARPES data didn't detect superconducting gaps  on the hole barrels, but did detect the gaps on the three $\Gamma-$ centered hole pockets which we consider.
 Several earlier works~\cite{Abdel-Hafiez2012,Fukazawa2009} analyzed the behavior of $C(T)$ in KFe$_2$As$_2$ within the phenomenological two-gap model, constructed in analogy with the two-gap model for MgB$_2$ (Ref. \cite{Golubov2002}). Our reasoning is similar to these works in the sense that we have larger gaps on $d_{xz}/d_{yz}$ pockets and a smaller gap on $D_{xy}$ pocket.  On the other hand, our analysis is based  microscopic three-band model, and we reproduce experimental $C(T)$
 with no free parameters.    

{\it \bf The model.}~~~The electronic structure of $\mathrm{KFe_2 As_2}$ in the physical 2-Fe Brillouin zone consists of 3 hole
pockets, located at the $\Gamma$-point, and hole barrels near $(\pi, \pi)$.  There is no evidence of superconductivity on the hole barrels, and we neglect them in our analysis.  Two inner $\Gamma$-centered pockets are made out of fermions from  $d_{xz}$ and $d_{yz}$ orbitals, and the outer pocket is made out  of fermions from $d_{xy}$ orbital \cite{okazaki2012}.
  We take as an input that the $d_{xy}$ band  has larger band mass/DOS than  $d_{xz}/d_{yz}$ bands.
We follow earlier works \cite{Cvetkovic2013,Vafek2017,Maiti2012,Classen2017}
and describe superconductivity within the low-energy model with $H=H_0+H_{int}$, where
 the quadratic Hamiltonian $H_0$ is given by $2 \times 2$ matrix for $d_{xz}$ and a separate term for $d_{yz}$ fermions, and $H_{int}$ is the Hubbard-Hund interaction, dressed by contributions from high-energy fermions.

To study superconductivity, we convert from orbital to band basis, i.e., diagonalize the quadratic form to $H_0 = \sum_k \varepsilon_{c,k} c^\dagger_k c_k + \varepsilon_{d,k} d^\dagger_k d_k + \varepsilon_{f,k} f^\dagger_k f_k$, where $c_k$ and $d_k$ are linear combinations of fermions from $d_{xz}$ and $d_{yz}$ orbitals, and $f$-operators describe $d_{xy}$ fermions.
The pairing interaction has $s$-wave and $d$-wave components (see Ref. \cite{Vafek2017} and Supplementary material (SM) for details).  We focus first on $s$-wave superconductivity  and discuss $d-$wave pairing later. The pairing interaction in $s-$wave channel is
\begin{equation}
\begin{gathered}
H_{SC}= \sum_{k,p, s\neq s'} \biggr[ U_{cc} c_{s k}^{\dagger} c_{s' -k}^{\dagger} c_{s' p} c_{s -p} + U_{dd} d_{s k}^{\dagger} d_{s' -k}^{\dagger} d_{s' p} d_{s -p} + \\ + U_{cd} \left( c_{s k}^{\dagger} c_{s' -k}^{\dagger} d_{s' p} d_{s -p}  + \mathrm{H.c.} \right) +
 U_{ff} f_{s k}^{\dagger} f_{s' -k}^{\dagger} f_{s' p} f_{s -p} \\ +
  \left(U_{fc} c_{s k}^{\dagger} c_{s' -k}^{\dagger} f_{s' p} f_{s -p} + U_{fd} d_{s k}^{\dagger} d_{s' -k}^{\dagger} f_{s' p} f_{s -p} + \mathrm{H.c.} \right) \biggr],
\end{gathered}
\label{band_H_BCS}
\end{equation}
where for circular hole pockets, bare interactions are
 $U_{cc} = U_{dd} = U_{cd} = (U+J')/2$, $U_{ff} = U/2$, and $U_{fc} = U_{fd} = \frac{J'}{2}$. After renormalizations  from high-energy fermions, all couplings become different, and, most important, $U^2_{cd}$ becomes larger than $U_{cc} U_{dd}$ (Refs.  \cite{Vafek2017, Chubukov2016,Maiti2012}). This gives rise to an attraction in the $s^{+-}$  channel.

\begin{figure}[h]
\center{\includegraphics[width=0.99\linewidth]{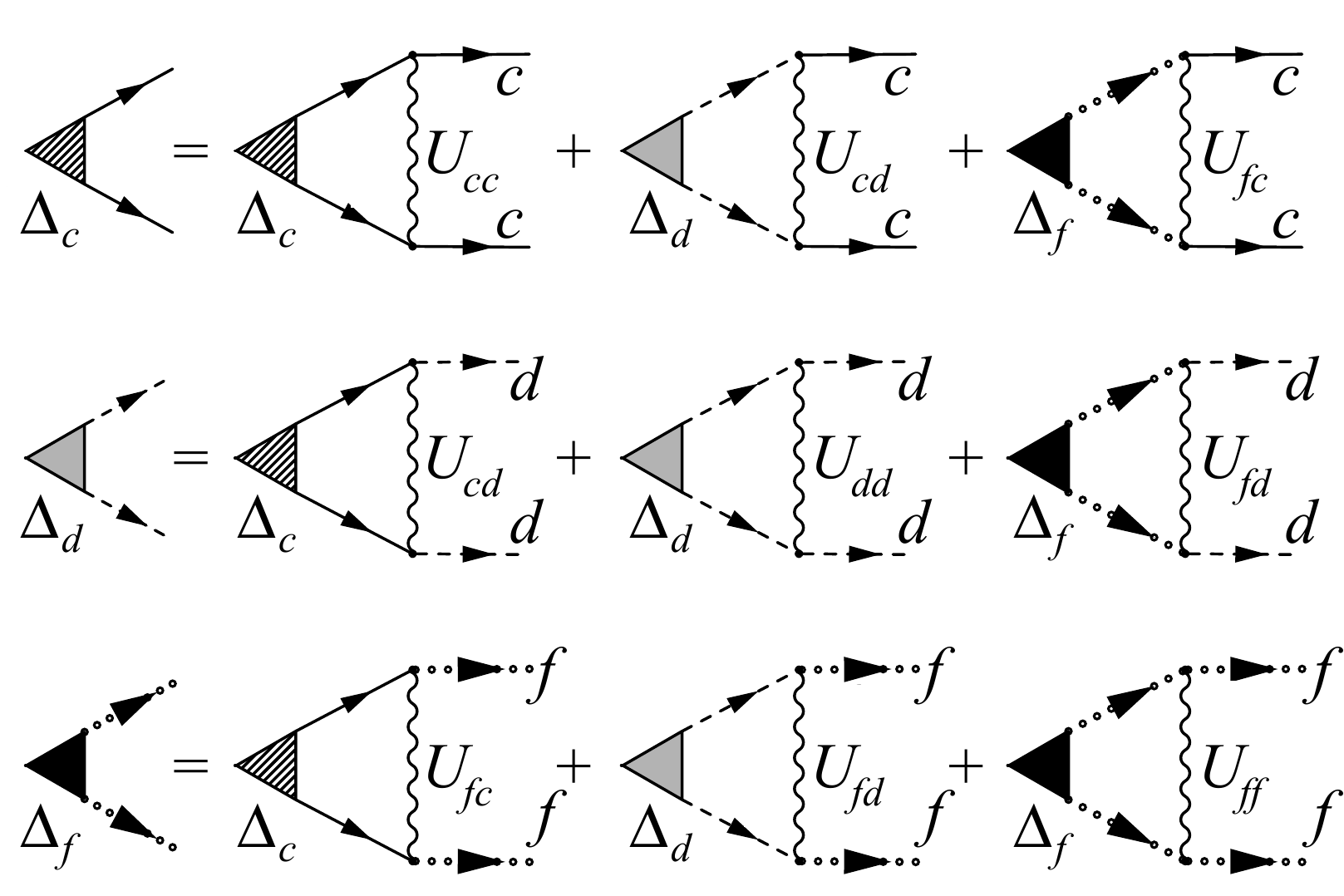}}
\centering{}\caption{Diagrammatic expressions for Gorkov's gap equations. Triangles with different filling represent SC vertexes on different bands. Solid, dashed, and dotted lines represent $c,d,f$-fermions respectively. Wavy lines represent interactions between fermions.}
\label{gap_diagr}
\end{figure}

{\it \bf Superconductivity.}~~~ Superconducting $T_c$ and $s-$wave  gaps on the three $\Gamma-$centered hole pockets at $T \leq T_c$ are obtained by solving the set of coupled linearized gap equations, presented in Fig. \ref{gap_diagr}.  In  analytical form we have
\begin{equation}
\begin{pmatrix}
\Delta_c \\
\Delta_d \\
\Delta_f
\end{pmatrix}= - L
\begin{pmatrix}
\nu_c U_{cc} & \nu_d U_{cd} & \nu_f U_{fc} \\
\nu_c U_{cd} & \nu_d U_{dd} & \nu_f U_{fd} \\
\nu_c U_{fc} & \nu_d U_{fd} & \nu_f U_{ff}
\end{pmatrix}
\begin{pmatrix}
\Delta_c \\
\Delta_d \\
\Delta_f
\end{pmatrix},
\label{gap_matr}
\end{equation}
where $L = \mathrm{ln} \frac{\Lambda}{T_c}$,  $\Lambda$ - is the upper cutoff, and $\nu_c, \nu_d,$ and $\nu_f$  are densities of states, proportional to the band masses.  In our case, $\nu_c \sim \nu_d$, and $\nu_f$ is larger.  We present the full solution for the gap in the SM and here show the
 result  for ${\vec \Delta} = (\Delta_c, \Delta_d, \Delta_f)^T =  \left(1, \alpha, -\beta \frac{\nu_c}{\nu_f}\right)^T$   to leading order in $\nu_{c,d}/\nu_f$,
where  $\alpha$ and $\beta$ do not depend on $\nu_f$ (see SM for exact expressions).
 The key observation here is that the gap $\Delta_f$ on the $d_{xy}$ pocket is small in the ratio of $\nu_{c,d}/\nu_f$.  This is the consequence of the fact that $s^{+-}$ superconductivity develops  on  $c$ and $d$ pockets (not to be confused with $s^{+-}$ pairing in systems with both electron and hole pockets), while the gap  on the $d_{xy}$ pocket does not develop on its own, but rather is  induced by  inter-orbital pairing interactions ($\Delta_f$ scales with $U_{fc}, U_{fd}$).   Note that $\Delta_f$ is non-zero only when $c$ and $d$ pockets are treated as non-equivalent, otherwise $\alpha =-1$ and $\beta =0$.

   To minimize the number of parameters, below we set $U_{cd}, U_{ff}, U_{fc}, U_{fd}$ equal to  their bare values in the Hubbard-Hund model (see above) and use $J=J'=0.4 U$ \cite{Classen2017}. Then $U_{cd} = 0.7U, U_{ff}=0.5U, U_{fc} = U_{fd} = 0.2U$ \cite{Bascones_review, Georges2013}.
 We model the renormalization of $U^2_{cd} - U_{cc} U_{dd}$ into a positive variable, necessary for $s^{+-}$ superconductivity,  by a single parameter $x$, by setting $U_{cc}= U_{cc}^{bare} (1-x) = 0.7 U(1-x), U_{dd} =  U_{dd}^{bare} (1+x) = 0.7U (1+x)$.
We used the experimental values $\nu_d / \nu_c=1.33, \nu_f / \nu_c=3.17$ from Ref.~\cite{Eilers2016} and set $x=0.5$ to match the experimental value of
$\alpha \approx -0.4$ (Ref. \cite{okazaki2012}). The same $x$ gives $\beta \nu_c/\nu_f \sim 0.06$, consistent with~\cite{okazaki2012}.

 {\it \bf The specific heat.}~~~
To calculate $C(T)$, we compute the internal energy $E(T)$ above and below $T_c$ and use $C(T) = dE/dT$.
To obtain $E(T)$  we construct a BCS Hamiltonian with anomalous terms with prefactors $\Delta_c$, $\Delta_d$, and $\Delta_f$, and diagonalize it. This yields
\begin{equation}
E(T) =- \sum_{i=c,d,f} \nu_i \int d\varepsilon_i \frac{\varepsilon_i^2 + \Delta_i^2 /2}{\sqrt{\varepsilon_i^2 + \Delta_i^2}} \tanh \frac{\sqrt{\varepsilon_i^2 + \Delta_i^2}}{2 T} + ...
\label{ex:aa}
\end{equation}
 where dots stand for temperature-independent terms.  We express $\Delta_d$ and $\Delta_f$ via $\Delta_c$  and
 $E(T)$ in powers of $\Delta_{c}$. To first order in $\nu_{c,d}/\nu_f$ we obtain
$ E (T)  = E (T_c)  -(\nu_c + \nu_d \alpha^2) |\Delta_c|^2/2$.
 The contribution from the $f$ band is small in $\nu_{c,d}/\nu_f$  despite that the  DOS  for this band is large. Using  $\Delta_c (T) \propto \sqrt{T_c - T}$, we then obtain that  the magnitude of the jump of the specific heat at $T_c$  does not depend on $\nu_f$.  The specific heat above $T_c$, on the other hand, comes primarily from the $d_{xy}$ band simply because DOS for this band is the largest.
 As a result, $\delta C/C_n  \propto \nu_{c,d}/\nu_f$ is small, unlike in a one-band BCS superconductor, where it is $O(1)$.
 We present the full expression for $\delta C/C_n$ in the SM.

 To obtain $C(T)$ below $T_c$, we
   assume, following \cite{Graser2009}
    that the ratios $\Delta_d/\Delta_c$ and $\Delta_f/\Delta_c$ remain the same as near $T_c$, and $\Delta_c (T)$ has the same
   temperature dependence as in BCS superconductor.
     We then find from (\ref{ex:aa}) that in the $T$ range where $\Delta_f (T) \approx -
     (\beta \nu_c/ \nu_f) \Delta_c (T)$ is smaller than $T$,  the contribution to the specific heat from the $d_{xy}$ band remains the same as in the normal state.  As the consequence,  $C(T)/T$ evolves from its maximal value right below  $T_c$ to a finite value equal  to the specific heat coefficient from non-superconducting $d_{xy}$ band.  This behavior changes below $T \sim T_{xy}$, at which $\Delta_f (T_{xy}) = T_{xy}$.  At such low temperatures the gap on the $d_{xy}$ band cannot be neglected, and the contribution to the specific heat from this band rapidly drops, and, as a result, $C(T)/T$ rapidly drops towards zero value at $T=0$.

\begin{figure}[h]
\center{\includegraphics[width=0.8\linewidth]{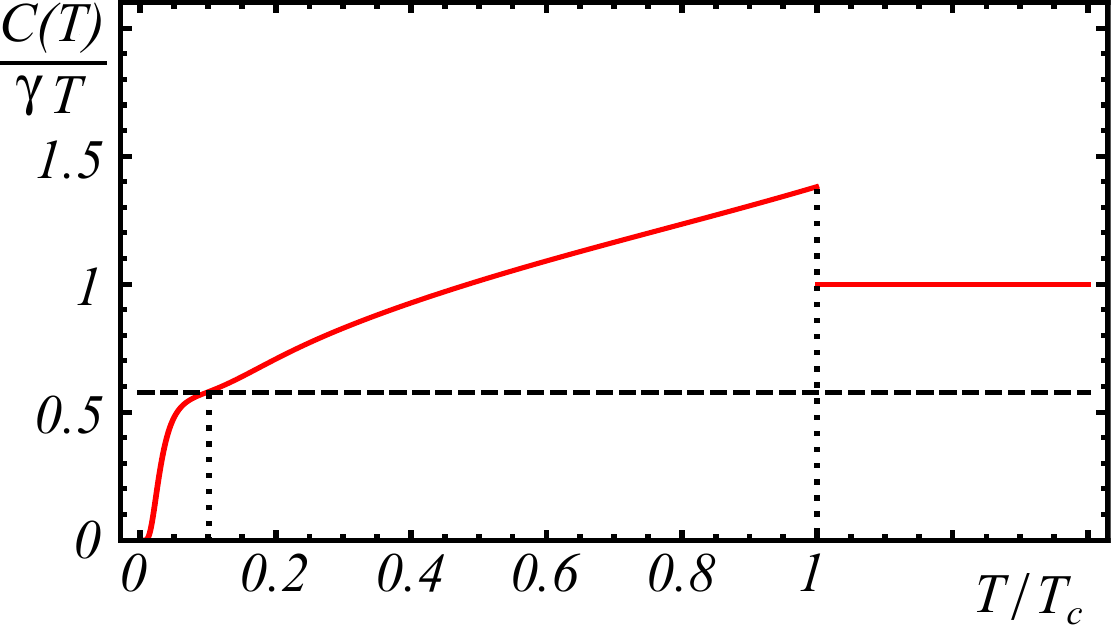}}
\centering{}\caption{The result of the numerical evaluation of the specific heat coefficient $C(T)/(\gamma T)$ within our model
($C(T) = \gamma T$ above $T_c$).
 The dashed line shows $C(T)/(\gamma T)$ for $d_{xy}$ orbital in the normal state.
The magnitude of the jump of $C(T)$ at $T_c$ and the overall  behavior of $C(T)/(\gamma T)$ below $T_c$   agrees well with the experimental data from \cite{Hardy2014,Abdel-Hafiez2012,Fukazawa2009,Eilers2016,Storey2013,Kim2011,Grinenko2014}.
 }
\label{heatcalc}
\end{figure}

 In Fig. \ref{heatcalc} we show the result of numerical calculation of the specific heat coefficient, using
  experimental values from the DOS's from Ref. ~\cite{Eilers2016}. The behavior is the same as presented schematically in  panel (c) of Fig. \ref{cplot}, and agrees {\it quantitatively} with the experimental data for KFe$_2$As$_2$ (Refs.  \cite{Hardy2014,Abdel-Hafiez2012,Fukazawa2009,Eilers2016,Storey2013,Kim2011,Grinenko2014}).
We emphasize that we fixed the only interaction parameter $x$ by matching the measured~\cite{okazaki2012} ratio of $\Delta_d/ \Delta_c$, hence our $C(T)$ is obtained with no fitting parameters. We reproduce  the experimental  location of $T_{xy}$, and the overall behavior of $C(T)$ below $T_c$.

 These results hold for circular hole pockets. For $C_4$ -symmetric, but non-circular pockets,
 $s-$wave components of the interactions and the gaps generally have additional $\cos{4n \theta}$ angular dependencies, which may give rise to accidental nodes~\cite{Maiti2012}.  This does not change much the  behavior of the specific heat at $T \leq T_c$, and only affects the functional form of $C(T)/T$ at the lowest $T$ where it drops anyway.

{\it \bf $d-$wave pairing.}~~~
Some experimental data, most notably on the  thermal conductivity \cite{Tafti2013,Abdel-Hafiez2013},  have been interpreted as evidence for $d-$wave pairing symmetry in KFe$_2$As$_2$. This is in variance with laser ARPES study \cite{okazaki2012,Ota2014}, whose results were interpreted as evidence for the $s-$wave pairing. Theoretical results show that $s-$wave and $d-$wave pairing components are both attractive and comparable in strength, with RPA calculations  \cite{Maiti2012} favoring $s^{+-}$ superconductivity and  early functional RG calculations \cite{Thomale2011} favoring $d-$wave pairing.  By all these reasons, it is instructive to analyze $C(T)$ for $d-$wave pairing.

 Within our model of circular pockets, $d-$wave pairing involves only $c$ and $d-$ pockets.  The $d-$wave component of the pairing interaction is
 \begin{equation}
\begin{gathered}
H_{SC}= \sum_{k,p, s\neq s'} \biggr[ {\tilde U}_{cc} c_{s k}^{\dagger} c_{s' -k}^{\dagger} c_{s' p} c_{s -p} + {\tilde U}_{dd} d_{s k}^{\dagger} d_{s' -k}^{\dagger} d_{s' p} d_{s -p} - \\ - {\tilde U}_{cd} \left( c_{s k}^{\dagger} c_{s' -k}^{\dagger} d_{s' p} d_{s -p}  + \mathrm{H.c.} \right) \biggr] \cos{2 \phi_k} \cos{2\phi_p},
\end{gathered}
\label{band_H_BCS_d}
\end{equation}
where $\phi_k$ and $\phi_p$ are angles along the Fermi surfaces. At the bare level
(i.e., without integrating out high-energy fermions)  ${\tilde U}_{cc} = {\tilde U}_{dd} = {\tilde U}_{cd} = (U-J')/2$.    There also exists the $\sin{2 \phi_k} \sin{2\phi_p}$ interaction component, but it does not give rise to new physics and we skip it.  After renormalization ${\tilde U}_{cc}$ and ${\tilde U}_{dd}$ split, and, most importantly, ${\tilde U}^2_{cd}$ becomes larger than ${\tilde U}_{cc} {\tilde U}_{dd}$ \cite{Xing2018Rapid}.
 Like for the $s^{+-}$ case, the enhancement of the inter-pocket pairing interaction gives rise to an attraction and a non-zero $T_c$ for $d-$wave pairing.
The matrix equation for the $d-$wave gap is
\begin{equation}
\begin{pmatrix}
\Delta_c \\
\Delta_d
\end{pmatrix}=-
\frac{L}{2}\begin{pmatrix}
\nu_c \tilde{U}_{cc} & -\nu_d \tilde{U}_{cd}  \\
-\nu_c \tilde{U}_{cd} & \nu_d \tilde{U}_{dd}
\end{pmatrix}
\begin{pmatrix}
\Delta_c \\
\Delta_d
\end{pmatrix}.
\end{equation}
Evaluating the eigenfunctions, substituting them into
 the expression for the internal energy $E(T)$, and differentiating over $T$, we obtain the behavior as in panel (d) of Fig. \ref{cplot}. Namely, the jump  $\delta C/C_n$ at $T_c$ is small, and $C(T)/T$ below $T_c$  drops but tends to a finite value at $T=0$, equal to $C(T)/T$ for non-superconducting $d_{xy}$ band.  This does not agree with the data, which clearly show that $C(T)/T$ drops below $T_{xy} <T_c$.
  This result holds for arbitrary $C_4$-symmetric dispersion, as long as the interaction  in the orbital basis is local, 
  and the larger hole pocket can be approximated as pure $d_{xy}$.  By all accounts (see e.g., Ref. \cite{Graser2009}), the admixture of $d_{xz}/d_{yz}$ orbital states to the composition of this pocket is very small (a percent), so $T_{xy}$, even if finite, should be truly small.
 
 {\it \bf Conclusions.}~~~
In this paper we studied the specific heat of
 $\mathrm{K Fe_2 As_2}$.
   We argued that $C(T)$ in the normal state is chiefly determined by the heavy $d_{xy}$ pocket, however superconductivity predominantly involves $d_{xz}/d_{yz}$ pockets, while the gap on the $d_{xy}$ pocket is either induced, but is small (for $s-$wave pairing), or not induced at all (for $d-$wave pairing). This gives rise to the behavior when (i) the jump of $C(T)$ at $T_c$ is much smaller than the BCS value,  and (ii) below $T_c$ specific heat
  coefficient $C(T)/T$ initially evolves towards a finite value, equal to normal state contribution  from $d_{xy}$ band.  For $s-$wave pairing, $C(T)/T$ eventually drops below a certain $T_{xy}$ (Figs. \ref{cplot} (c)  and \ref{heatcalc}). If the pairing is $d-$wave, $T_{xy} =0$ in our analysis, and  is likely quite small in a more general case. The experimentally detected behavior of $C(T)/T$ (Refs. \cite{Hardy2014,Abdel-Hafiez2012,Fukazawa2009,Eilers2016,Storey2013,Kim2011})
  is more consistent  with $s-$wave pairing.
We used the detuning of interactions on $d_{xz}$ and $d_{yz}$ pockets from their bare values as a single adjustable parameter to reproduce the data on gap ratio on the two small pockets~\cite{okazaki2012}. After that,  our theory has no free parameters. It reproduces the magnitude of the jump at $T_c$, the shape of $C(T)/T$ below $T_c$, and the value of $T_{xy}$. We emphasize  that we did not assume that interactions involving $d_{xy}$ fermions are additionally  reduced due to potentially small quasiparticle residue $Z$ for fermions on the $d_{xy}$ band.
   The reduction of $Z_{xy}$  under hole doping follows from quite solid theoretical arguments~\cite{deMedici2014,pavarini2017}, what is less clear is whether the reduction is strong enough to affect $C(T)$. If it is,  the overall behavior of $C(T)/T$ will not change compared to our analysis, but $T_{xy}$  will decrease further compared to $T_c$.  A systematic study of $C(T)$ in  doped  K$_{1-x}$Ba$_x$Fe$_2$As$_2$ is needed to determine the influence of $Z_{xy}$ on the specific heat.

\begin{acknowledgments}
 We thank  R. Fernandes, F. Hardy, H. v L\"{o}hneysen, D. Shaffer, and R-Q Xing for useful discussions. The work was supported  by the Office
of Basic Energy Sciences, U.S. Department of Energy, under award DE-SC0014402 (AVC).
\end{acknowledgments}

\bibliography{biblio}

\end{document}


\title{Supplementary materials for "Specific heat in strongly hole-doped Iron-based  superconductors"}

\author{Dmitry V. Chichinadze and Andrey V. Chubukov}
\affiliation{School of Physics and Astronomy,
University of Minnesota, Minneapolis, MN 55455, USA}

\maketitle

\section{The microscopic low-energy model}
\label{sec:sys}

\begin{figure}[h]
\center{\includegraphics[width=0.75\linewidth]{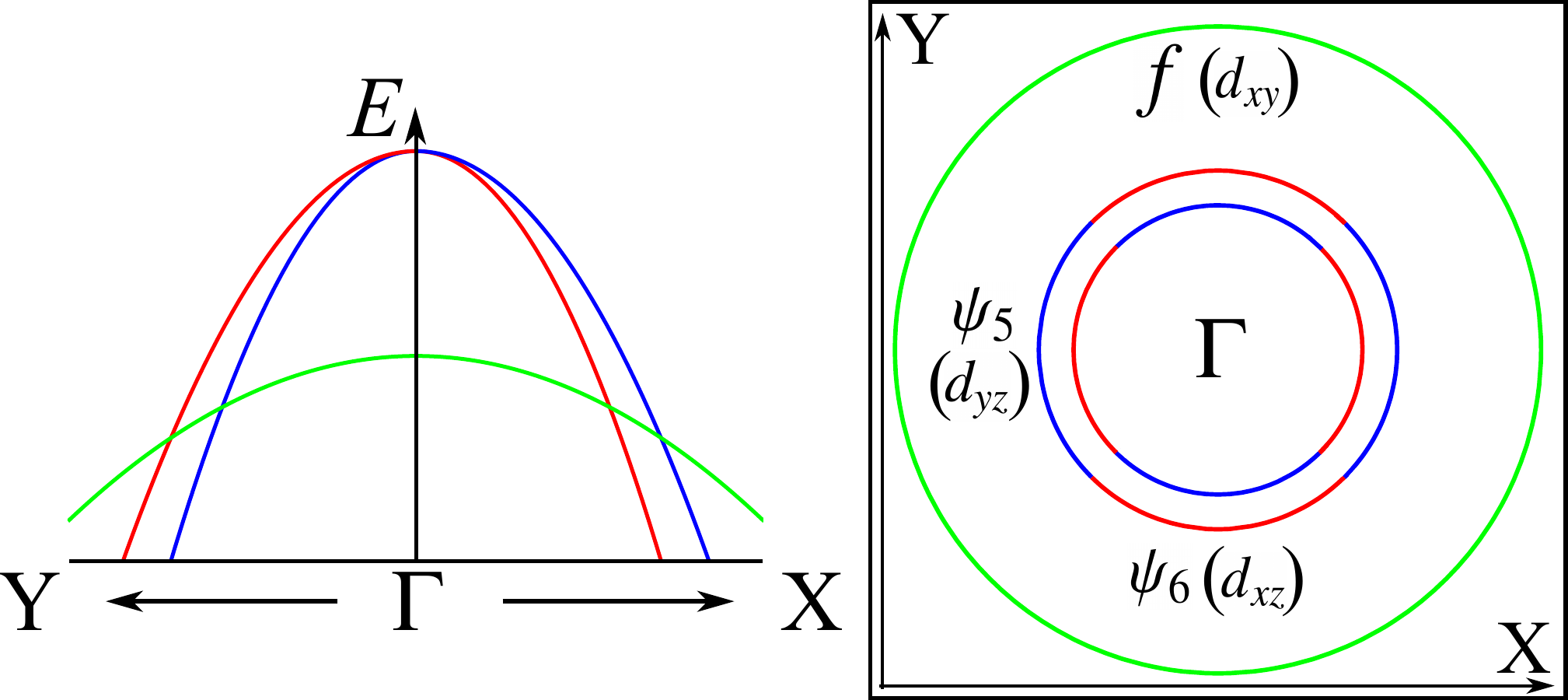}}
\centering{}\caption{Left panel: energy spectrum of our model near the $\Gamma$-point in 1 Fe Brillouin zone. This spectrum reproduces the electronic
spectrum obtained from first-principles calculations in \cite{okazaki2012}. Right panel: sketch of three pockets in the Brillouin zone. The green line represents $d_{xy}$-orbital pocket, the blue line represents $d_{yz}$, and the red line represents $d_{xz}$-orbital pocket.}
\label{sk}
\end{figure}

In this paper we use the approach introduced by Vafek and Cvetkovic \cite{Cvetkovic2013, Vafek2017, Chubukov2016},
where the quadratic Hamiltonian $H_0$ is given by the matrix product of low-energy spinor wave function $\Psi^{\dagger}_{k,s} = \left( \psi^{\dagger}_{5k,s}, \psi^{\dagger}_{6k,s}, f^{\dagger}_{k,s}  \right)$  (we use the same notations as in Ref. \cite{Xing2017})
\begin{equation}
H_0 = \sum_{\vec{k}, s}
\begin{pmatrix}
\psi_{5k}^{\dagger} \\
\psi_{6k}^{\dagger} \\
f_k^{\dagger}
\end{pmatrix}^T
\begin{pmatrix}
\mu - \frac{\vec{k}^2}{2 m} + b \left( k_y^2 - k_x^2 \right) && 2 c k_x k_y && 0 \\
2 c k_x k_y && \mu - \frac{\vec{k}^2}{2 m} - b \left( k_y^2 - k_x^2 \right) && 0 \\
0 && 0 && \mu - \frac{\vec{k}^2}{2 M}
\end{pmatrix}
\begin{pmatrix}
\psi_{5k} \\
\psi_{6k} \\
f_k
\end{pmatrix}.
\label{h0}
\end{equation}
Here operator $\psi_5$ represents electrons on the $d_{yz}$-orbital, $\psi_6$ represents $d_{xz}$-orbital, while $f$ represents $d_{xy}$-orbital and the sum is taken over momenta around the $\Gamma$-point and both spin projections (we omit  spin indexes for simplicity).   The parameters $b, c$ define the shape of the FS and are obtained by fitting the ARPES data \cite{Xing2017}. In this paper we neglect $C_4$-symmetric variation of Fermi momentum along the hole pockets
and consider them as circular. This corresponds to setting $b=c$ in Eq. (\ref{h0}) \cite{Xing2017}. The band mass $M$ of the $d_{xy}$-orbital is larger than the band mass $m$ of $d_{xz}/d_{yz}$, so $M>>m$.

Departing from the local Hubbard-Hund interactions one can cast \cite{Cvetkovic2013,Chubukov2016} the interaction part of the Hamiltonian in the form
\begin{equation}
\begin{split}
H_{int} = \frac{U}{2} \sum \left[ \psi_{5 s k}^{\dagger} \psi_{5 s' -k}^{\dagger} \psi_{5 s' p} \psi_{5 s -p} + \psi_{6 s k}^{\dagger} \psi_{6 s' -k}^{\dagger} \psi_{6 s' p} \psi_{6 s -p} \right] +
\frac{J'}{2} \sum \left[ \psi_{5 s k}^{\dagger} \psi_{5 s' -k}^{\dagger} \psi_{6 s' p} \psi_{6 s -p} + \psi_{6 s k}^{\dagger} \psi_{6 s' -k}^{\dagger} \psi_{5 s' p} \psi_{5 s -p} \right]\\
+ \frac{U'}{2} \sum \left[ \psi_{5 s k}^{\dagger} \psi_{6 s' -k}^{\dagger} \psi_{5 s' p} \psi_{6 s -p} + \psi_{6 s k}^{\dagger} \psi_{5 s' -k}^{\dagger} \psi_{6 s' p} \psi_{5 s -p} \right] +
\frac{J}{2} \sum \left[ \psi_{5 s k}^{\dagger} \psi_{6 s' -k}^{\dagger} \psi_{6 s' p} \psi_{5 s -p} + \psi_{6 s k}^{\dagger} \psi_{5 s' -k}^{\dagger} \psi_{5 s' p} \psi_{6 s -p} \right] + \\
\frac{U}{2} \sum \left[ f_{s k}^{\dagger} f_{s' -k}^{\dagger} f_{s' p} f_{s -p}  \right] + \frac{J'}{2} \sum \left[ \psi_{5 s k}^{\dagger} \psi_{5 s' -k}^{\dagger} f_{s' p} f_{s -p} + f_{s k}^{\dagger} f_{s' -k}^{\dagger} \psi_{5 s' p} \psi_{5 s -p} \right] +
\frac{J'}{2} \sum \left[ \psi_{6 s k}^{\dagger} \psi_{6 s' -k}^{\dagger} f_{s' p} f_{s -p} + f_{s k}^{\dagger} f_{s' -k}^{\dagger} \psi_{6 s' p} \psi_{6 s -p} \right],
\end{split}
\label{orb_H}
\end{equation}
where $U$ -- is the Hubbard intraorbital interaction, $U'$ -- is the Hubbard interorbital interaction, $J$ -- is the Hund's exchange interaction, and $J'$ -- is the
amplitude of the interorbital pair hopping \cite{Chubukov2016, Xing2017}.  Interaction terms like $\psi_{5 s k}^{\dagger} \psi_{6 s' -k}^{\dagger} f_{s' p} f_{s -p}$ are not included in Eq. (\ref{orb_H}) because they do not influence superconductivity.

We now perform the standard rotation transformation, which changes representation from the orbital to the band one. This transformation is applicable only to $\psi$-electrons, because $f$-electrons are
already in the diagonal basis. The transformation is given by the matrix equation \cite{Xing2017, Chubukov2016}
\begin{equation}
\begin{pmatrix}
\psi_{5k} \\
\psi_{6k}
\end{pmatrix}=
\begin{pmatrix}
\mathrm{cos}\phi_k & \mathrm{sin}\phi_k\\
-\mathrm{sin}\phi_k & \mathrm{cos}\phi_k
\end{pmatrix}
\begin{pmatrix}
c_k \\
d_k
\end{pmatrix}.
\label{rot_trans}
\end{equation}
The non-interacting part $H_0$ of the Hamiltonian in diagonal variables $c,d$ and $f$ then reads
$H_0=\sum_{k,s} \left[ \varepsilon_c c_k^{\dagger} c_k + \varepsilon_d d_k^{\dagger} d_k + \varepsilon_f f_k^{\dagger} f_k \right],$ where the band masses $m_{c,d}$ depend on the FS shape parameters $b$ and $c$ \cite{Xing2017}, but are still smaller than the $f$-band ($d_{xy}$) mass $M$. In the system we consider we can omit inter-band pairing condensates $\av{c^{\dagger} d^{\dagger}}$ due to their smallness with respect to intra-band condensates $\av{c^{\dagger} c^{\dagger}}$, etc.
Applying transformation (\ref{rot_trans}) to the Hamiltonian (\ref{orb_H}) and only keeping terms which contribute to the SC channel we obtain the pairing Hamiltonian in the band basis
\begin{equation}
\begin{gathered}
H_{SC}= \sum \biggr[
W_{cc} c_{s k}^{\dagger} c_{s' -k}^{\dagger} c_{s' p} c_{s -p} + W_{dd} d_{s k}^{\dagger} d_{s' -k}^{\dagger} d_{s' p} d_{s -p} + W_{cd} \left( c_{s k}^{\dagger} c_{s' -k}^{\dagger} d_{s' p} d_{s -p} + d_{s k}^{\dagger} d_{s' -k}^{\dagger} c_{s' p} c_{s -p}  \right) + \\
U_{ff} f_{s k}^{\dagger} f_{s' -k}^{\dagger} f_{s' p} f_{s -p} + \left(U_{fc}  \right) \left(c_{s k}^{\dagger} c_{s' -k}^{\dagger} f_{s' p} f_{s -p} + f_{s k}^{\dagger} f_{s' -k}^{\dagger} c_{s' p} c_{s -p} \right) +
 \left(U_{fd}  \right) \left(d_{s k}^{\dagger} d_{s' -k}^{\dagger} f_{s' p} f_{s -p} + f_{s k}^{\dagger} f_{s' -k}^{\dagger} d_{s' p} d_{s -p} \right)  \biggr],
\end{gathered}
\label{band_H_BCS}
\end{equation}
where
\begin{equation}
\begin{gathered}
W_{cc,dd} = U_{cc,dd} + \tilde{U}_{cc,dd} \cos 2\phi_k \cos 2\phi_p +  \tilde{\tilde{U}}_{cc,dd} \mathrm{sin}2\phi_k \mathrm{sin}2\phi_p,\\
W_{cd} = U_{cd} - \tilde{U}_{cd} \cos 2\phi_k \cos 2\phi_p -  \tilde{\tilde{U}}_{cd} \mathrm{sin}2\phi_k \mathrm{sin}2\phi_p.
\end{gathered}
\end{equation}
At the bare level
\begin{equation}
U_{cc}=U_{dd}=U_{cd}=\frac{U+J'}{2}, \; \tilde{U}_{cc} = \tilde{U}_{dd} = \tilde{U}_{cd}=\frac{U-J'}{2}, \; \tilde{\tilde{U}}_{cc} = \tilde{\tilde{U}}_{dd} = \tilde{\tilde{U}}_{cd} =\frac{U'+J}{2}, \; U_{fc} = U_{fd} = \frac{J'}{2}, U_{ff} = \frac{U}{2}.
\label{coupl_bare}
\end{equation}

\section{Gap equation and the structure of the SC order parameter }
\label{sec:gapeq}

\begin{figure}[h]
\center{\includegraphics[width=0.65\linewidth]{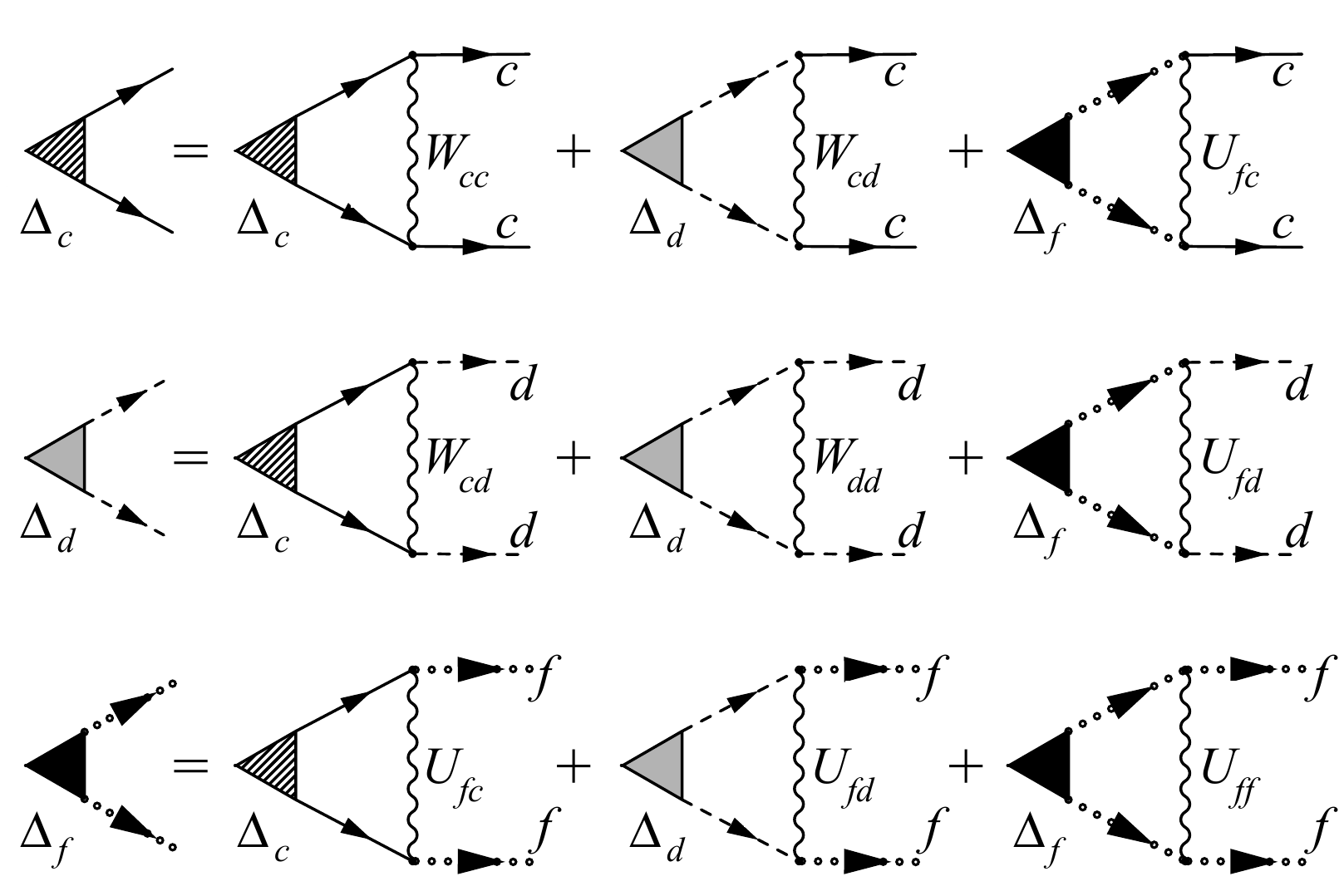}}
\centering{}\caption{Diagrammatic expressions for Gor'kov (gap) equations. Triangles with different filling represent SC vertexes on different bands. Solid, dashed, and dotted lines represent $c,d,f$-fermions respectively. Wavy lines represent interactions between fermions.}
\label{gap_diagr}
\end{figure}

To solve for superconductivity (SC) we write down the qap equations in the band basis using diagrammatic Gor'kov equations, presented in the main text.
Here we have introduced three superconducting gaps $\Delta_c, \Delta_d, \Delta_f$, so our order parameter is an $O(3)$ vector. We seek for a solution with preserved time reversal (TR) symmetry. Under this condition there cannot be phase difference between components of the $O(3)$ vector other then 0 or $\pi$. In the band basis representation orbital content only gives the angular dependence of the interactions, thus generating SC gaps with different symmetries. In the operator form the gap equation reads
\begin{equation}
\begin{split}
\Delta_c = -\left(W_{cc} \Pi_{cc} \Delta_c + W_{cd} \Pi_{dd} \Delta_d + U_{fc} \Pi_{ff} \Delta_f \right), \\
\Delta_d = -\left(W_{cd} \Pi_{cc} \Delta_c + W_{dd} \Pi_{dd} \Delta_d + U_{fd} \Pi_{ff} \Delta_f \right), \\
\Delta_f = -\left(U_{fc} \Pi_{cc} \Delta_c + U_{fd} \Pi_{dd} \Delta_d + U_{ff} \Pi_{ff} \Delta_f \right), \\
\end{split}
\label{lin_gap}
\end{equation}
where $\Pi_{jj}$ - is the standard polarization bubble given by
\begin{equation}
\Pi_{jj} = \frac{i}{(2\pi)^2} (\delta_{\alpha \gamma} \delta_{\beta \delta} - \delta_{\alpha \delta} \delta_{\beta \gamma}) \int d^2 q \frac{\mathrm{th}\frac{\varepsilon^j_q}{2T}}{2 \varepsilon^j_q},
\end{equation}
with $j=c,d,f$ for different electron flavors. Within our model $s-$ and $d-$wave channels do factorize. One can show this explicitly starting from integration over $k$ in polar coordinates to obtain for the linearized equation:
\begin{equation}
\Pi_{jj} \propto \int d^2 q \frac{\mathrm{th}\frac{\varepsilon^j_q}{2T}}{2 \varepsilon^j_q} = \nu_j \int d\varepsilon d\theta \frac{\mathrm{th}\frac{\varepsilon}{2T}}{2 \varepsilon} = \frac{\nu_j}{2} \mathrm{ln} \frac{\Lambda}{T} \int d\theta ...,
\end{equation}
where $\Lambda$ - is the cutoff frequency and $\nu_j$ - is the $j$-electron DOS at Fermi energy. In 2D DOS $\nu_j=m_j / (2 \pi)$, where $m_f=M$ and $m_c, m_d$ are expressed via parameters of $H_0$ \cite{Xing2017}: $m_{c,d} = m(1 \mp 2m b)$. 
In order to solve the gap equation it is convenient to introduce gap functions averaged over the angle in the BZ:
\begin{equation}
\bar{\Delta}_{c,d,f}^{(1)} = \int d\theta \, \Delta_{c,d,f}(\theta), \; \bar{\Delta}_{c,d}^{(2)} = \int d\theta \, \Delta_{c,d}(\theta) \cos 2\theta, \; \bar{\Delta}_{c,d}^{(3)} = \int d\theta \, \Delta_{c,d}(\theta) \sin 2\theta.
\end{equation}
Using these quantities we can rewrite Eq. (\ref{lin_gap}):
\begin{equation}
\begin{gathered}
\bar{\Delta}_c^{(1)} = -L\left(\nu_c U_{cc} \bar{\Delta}_c^{(1)} + \nu_d U_{cd} \bar{\Delta}_d^{(1)} + \nu_f U_{fc} \bar{\Delta}_f^{(1)}\right), \\
\bar{\Delta}_c^{(2)} = -L\left(\frac{\nu_c}{2} \tilde{U}_{cc} \bar{\Delta}_c^{(2)} - \frac{\nu_d}{2} \tilde{U}_{cd} \bar{\Delta}_d^{(2)}\right), \\
\bar{\Delta}_c^{(3)} = -L\left(\frac{\nu_c}{2} \tilde{\tilde{U}}_{cc} \bar{\Delta}_c^{(3)} - \frac{\nu_d}{2} \tilde{\tilde{U}}_{cd} \bar{\Delta}_d^{(3)}\right), \\
\bar{\Delta}_d^{(1)} = -L\left(\nu_c U_{cd} \bar{\Delta}_c^{(1)} + \nu_d U_{dd} \bar{\Delta}_d^{(1)} + \nu_f U_{fd} \bar{\Delta}_f^{(1)}\right), \\
\bar{\Delta}_d^{(2)} = -L\left(-\frac{\nu_c}{2} \tilde{U}_{cd} \bar{\Delta}_c^{(2)} + \frac{\nu_d}{2} \tilde{U}_{dd} \bar{\Delta}_d^{(2)}\right), \\
\bar{\Delta}_d^{(3)} = -L\left(-\frac{\nu_c}{2} \tilde{\tilde{U}}_{cd} \bar{\Delta}_c^{(3)} + \frac{\nu_d}{2} \tilde{\tilde{U}}_{dd} \bar{\Delta}_d^{(3)}\right), \\
\bar{\Delta}_f^{(1)} = -L\left(\nu_c U_{fc} \bar{\Delta}_c^{(1)} + \nu_d U_{fd} \bar{\Delta}_d^{(1)} + \nu_f U_{ff} \bar{\Delta}_f^{(1)}\right),
\end{gathered}
\label{gap_alg}
\end{equation}
where $L = \mathrm{ln} \frac{\Lambda}{T_c}$. According to the Eq. (\ref{gap_alg}) $\bar{\Delta}_c^{(1)}=\bar{\Delta}_d^{(1)}$,
$\bar{\Delta}_c^{(2)}=-\bar{\Delta}_d^{(2)}$, $\bar{\Delta}_c^{(3)}=-\bar{\Delta}_d^{(3)}$ for bare couplings. Thus, there exist only 4 linearly-independent components, and both $s$- and $d$-wave channels are decoupled. In other words, for both SC channels there exist zero eigenvalues of the gap equation (\ref{gap_alg}). Therefore, the system can exhibit SC if renormalized interactions instead of bare are taken into account \cite{Vafek2017}. This property arises from an instability of zero eigenvalues of gap equation with bare interactions. In the standard BCS theory infinitesimally small attraction in the Cooper channel, which is mediated by the electron-phonon interaction, already results in an instability. Here the attraction is of the Kohn-Luttinger origin, i.e. it comes from the renormalization of repulsive Coulomb interaction. Below we consider renormalization and solve the gap equation for both $s$- and $d$-wave channels.

\subsection{$s$-wave solution}
\label{sec:swave}

Let us first look for the $s$-wave solution. Both $s$- and $d-$channels decouple in Eq. (\ref{gap_alg}), thus the $s-$wave gap equation (\ref{gap_alg}) can be written in the $3\times 3$ matrix form
\begin{equation}
\begin{pmatrix}
\bar{\Delta}_c^{(1)} \\
\bar{\Delta}_d^{(1)} \\
\bar{\Delta}_f
\end{pmatrix}= -L
\begin{pmatrix}
\nu_c U_{cc} & \nu_d U_{cd} & \nu_f U_{fc} \\
\nu_c U_{cd} & \nu_d U_{dd} & \nu_f U_{fd} \\
\nu_c U_{fc} & \nu_d U_{fd} & \nu_f U_{ff}
\end{pmatrix}
\begin{pmatrix}
\bar{\Delta}_c^{(1)} \\
\bar{\Delta}_d^{(1)} \\
\bar{\Delta}_f
\end{pmatrix},
\label{gap_eq_tempsup}
\end{equation}
where all the interactions are bare, see Eq. (\ref{coupl_bare}). Previous studies \cite{Vafek2017} have shown, that in the low-energy theory one has to consider the gap equation with the renormalized couplings instead of bare interactions. This can be seen as a consequence of renormalization in either orbital, or band basis (see Supplementary materials to \cite{Vafek2017}). One can build up such low-energy theories in two ways. The first way is to start from the orbital basis. Then one gets identical interactions for $c$ and $d$ bands as in Eq. (\ref{gap_eq_tempsup}), but this model does
not allow for SC to occur, so one needs to renormalize interactions. Within renormalization group (RG) interactions change differently, thus they are not identical for different bands anymore, and one now can solve for SC. The other way is to start from the band basis
with different interactions for $c$ and $d$ electrons, so that the SC is possible from the beginning. Doing the "inverse" transformation to the orbital basis one will notice, that the inequality of interactions in the band basis results in extra terms in the
Hamiltonian written in the band basis.

Here we use the first approach. Within RG, intra-band couplings $W_{cc}, W_{dd},$ and $W_{cd}$ are being renormalized differently \cite{Vafek2017}, which allows for SC to occur. In the general case one has to solve the Eq. (\ref{gap_alg}) with renormalized couplings, which then will be a $7 \times 7$ matrix equation with decoupling channels.  Here we consider the renormalization effects in the $s-$wave channel, thus the gap equation ($\ref{gap_eq_tempsup}$) is now written with renormalized interactions, i.e. $U_{cc} \neq U_{dd} \neq U_{cd}$ (or at least $U_{cc,dd} \neq U_{cd},$ see Supplementary materials to \cite{Vafek2017}).
Although Eq. (\ref{gap_eq_tempsup}) is solvable analytically, its solution is not informative because it is too cumbersome, so we will solve the gap equation (\ref{gap_eq_tempsup}) perturbatively in small $\nu_c, \nu_d / \nu_f \propto m/M$ instead.

We start from the equation for $\Delta_f$ (within this section we further use $\bar{\Delta}_c^{(1)} = \Delta_c, \bar{\Delta}_d^{(1)} =\Delta_d, \bar{\Delta}_f=\Delta_f $ for shortness):
\begin{equation}
\Delta_f = \frac{-L }{1 + L \nu_f U_{ff}} \left( \nu_c U_{fc} \Delta_c + \nu_d U_{fd} \Delta_d \right).
\label{del_gap_ex}
\end{equation}
We next substitute this relation to the equations for $\Delta_c, \Delta_d$
\begin{equation}
\begin{cases}
\Delta_c = -L \left[ \nu_c U_{cc} \Delta_c + \nu_d U_{cd} \Delta_d - \frac{L \nu_f U_{fc}}{1 + L \nu_f U_{ff}} \left( \nu_c U_{fc} \Delta_c + \nu_d U_{fd} \Delta_d \right) \right], \\
\Delta_d = -L \left[ \nu_c U_{cd} \Delta_c + \nu_d U_{dd} \Delta_d - \frac{L \nu_f U_{fd}}{1 + L \nu_f U_{ff}} \left( \nu_c U_{fc} \Delta_c + \nu_d U_{fd} \Delta_d \right)  \right].
\label{two_gap_alg}
\end{cases}
\end{equation}
For large $\nu_f$ within the leading order
\begin{equation}
\frac{L \nu_f U_{fc}}{1 + L \nu_f U_{ff}} = \frac{U_{fc}}{U_{ff}}, \; \frac{L \nu_f U_{fd}}{1 + L \nu_f U_{ff}} = \frac{U_{fd}}{U_{ff}}.
\label{lead_ord}
\end{equation}
Then equations for $\Delta_c, \Delta_d$ can be written as
\begin{equation}
\begin{pmatrix}
\Delta_c \\
\Delta_d
\end{pmatrix} = -L
\begin{pmatrix}
\nu_c \bar{U}_{cc} & \nu_d \bar{U}_{cd} \\
\nu_c \bar{U}_{cd} & \nu_d \bar{U}_{dd}
\end{pmatrix}
\begin{pmatrix}
\Delta_c \\
\Delta_d
\end{pmatrix},
\label{gapeq_pertsup}
\end{equation}
where 
$$
\bar{U}_{cc} = U_{cc} - \frac{U_{fc}^2}{U_{ff}}, \bar{U}_{dd} = U_{dd} - \frac{U_{fd}^2}{U_{ff}}, \bar{U}_{cd} = U_{cd} - \frac{U_{fc} U_{fd}}{U_{ff}}.
$$
The square of gap ratio $(\Delta_d / \Delta_c)^2$ can be directly obtained from Eqs. (\ref{two_gap_alg}), (\ref{lead_ord}):
\begin{equation}
\left( \frac{\Delta_d}{\Delta_c} \right)^2 = \frac{\nu_c}{\nu_d} \frac{(1+ L \nu_c \bar{U}_{cc})}{(1+ L \nu_d \bar{U}_{dd})}.
\label{gap_ratio}
\end{equation}
For bare couplings $\bar{U}_{cc}=\bar{U}_{dd}=\bar{U}_{cd}$, therefore the only eigenvalue of gap matrix which can give attraction  
\begin{equation}
\lambda = \frac{\bar{U}_{cc} \nu_c+ \bar{U}_{dd} \nu_d - \sqrt{(\bar{U}_{cc} \nu_c + \bar{U}_{dd} \nu_d)^2 + 4  \nu_c \nu_d (\bar{U}_{cd}^2 - \bar{U}_{cc} \bar{U}_{dd}) } }{2}
\end{equation}
is zero, thus one has to consider renormalization of interactions in order to get attraction in the channel. This renormalization can be obtained by transformation $U_{cc} \to U^{bare}_{cc} (1-x), U_{dd} \to U^{bare}_{dd} (1+x)$, with $0<x<1$ being a free parameter. Under this transformation  $\bar{U}_{cd}^2 > \bar{U}_{cc} \bar{U}_{dd}$ for any non-zero value of $x$, which allows for attraction to appear. From Eq. (\ref{gapeq_pertsup}) we also get $L$, which corresponds to largest eigenvalue:
\begin{equation}
L = \frac{-\nu_c \bar{U}_{cc} - \nu_d \bar{U}_{dd} - \sqrt{(\nu_c \bar{U}_{cc} + \nu_d \bar{U}_{dd})^2 - 4 \nu_c \nu_d \left( \bar{U}_{cc} \bar{U}_{dd} - \bar{U}_{cd}^2 \right)} }{2 \nu_c \nu_d \left( \bar{U}_{cc} \bar{U}_{dd} - \bar{U}_{cd}^2 \right)},
\end{equation}
and the solution of the gap equation for $\Delta_c, \Delta_d$ (setting $\Delta_c=1$): 
\begin{equation}
\Delta_c = 1, \Delta_d = \alpha,
\end{equation}
with
\begin{equation}
\alpha = \frac{(\nu_d \bar{U}_{dd} - \nu_c \bar{U}_{cc} ) - \sqrt{(\nu_d \bar{U}_{dd}  - \nu_c \bar{U}_{cc} )^2 + 4 \nu_c \nu_d \bar{U}_{cd}^2 }}{2  \nu_d \bar{U}_{cd} }.
\label{alpha_def}
\end{equation}

Finally, using Eq. (\ref{del_gap_ex}) we express $\Delta_f$ via $\Delta_c$
\begin{equation}
\Delta_f = -\frac{\nu_c}{\nu_f} \frac{L \nu_f}{1 + L \nu_f U_{ff}} \left( U_{fc} \Delta_c + \frac{\nu_d}{\nu_c} U_{fd} \Delta_d \right) =  -\frac{\nu_c}{\nu_f} \frac{L \nu_f}{1 + L \nu_f U_{ff}} \left( U_{fc} + \alpha \frac{\nu_d}{\nu_c} U_{fd} \right),
\end{equation}
so the full gap equation solution within the leading order reads
\begin{equation}
\vec{\Delta} = (1, \alpha, - \beta \frac{\nu_c}{\nu_f} ),
\label{gap_eq_solsup}
\end{equation}
where $\alpha$ is given by Eq. (\ref{alpha_def}) and
\begin{equation}
\beta = \frac{1}{U_{ff}} \left( U_{fc} + \alpha \frac{\nu_d}{\nu_c} U_{fd} \right).
\label{beta_def}
\end{equation}

\subsection{$d$-wave solution}

Instead of the $s$-wave one can look for a $d$-wave solution. Then according to Eq. (\ref{gap_alg}) and since the channels decouple
the gap equation reads
\begin{equation}
\begin{pmatrix}
\Delta_c \\
\Delta_d
\end{pmatrix}=-
\frac{L}{2}\begin{pmatrix}
\nu_c \tilde{U}_{cc} & -\nu_d \tilde{U}_{cd}  \\
-\nu_c \tilde{U}_{cd} & \nu_d \tilde{U}_{dd}
\end{pmatrix}
\begin{pmatrix}
\Delta_c \\
\Delta_d
\end{pmatrix}.
\end{equation}
For bare couplings one of the eigenvalues is again zero. This allows to find the solution if renormalized interactions are taken into account:
\begin{equation}
\vec{\Delta} = \left(-\frac{\nu_c {\tilde U}_{cc} - \nu_d {\tilde U}_{dd} - \sqrt{(\nu_c {\tilde U}_{cc} - \nu_d {\tilde U}_{dd})^2 + 4 \nu_c \nu_d {\tilde U}_{cd}^2}}{2 \nu_c {\tilde U}_{cd}}, 1 \right)^T.
\label{dwavesolsup}
\end{equation}
The solution (\ref{dwavesolsup}) with renormalized couplings allows for SC to occur.

\section{Calculations of the specific heat at and below $T_c$}
\label{sec:thermo}

To calculate the specific heat we first construct the mean-field BCS-like Hamiltonian made out of the Hamiltonian (\ref{band_H_BCS})
\begin{equation}
H_{MF} = \sum \varepsilon_c c_k^{\dagger} c_k + \varepsilon_d d_k^{\dagger} d_k + \varepsilon_f f_k^{\dagger} f_k + \Delta_c  c^{\dagger}_k c^{\dagger}_k + \Delta_d d^{\dagger}_k d^{\dagger}_k + \Delta_f f^{\dagger}_k f^{\dagger}_k + \Delta_i \hat{g}^{-1}_{ij} \Delta_j + \text{H.c.}
\label{HBdG}
\end{equation}
where we neglected the constant term $-\frac{1}{2} \sum_k (\varepsilon_c + \varepsilon_d + \varepsilon_f)$ because it doesn't contribute to the specific heat and
$$\hat{g}^{-1} = \begin{pmatrix}
U_{cc} & U_{cd} & U_{fc}\\
U_{cd} & U_{dd} & U_{fd}\\
U_{fc} & U_{fd} &  U_{ff}
\end{pmatrix}^{-1}$$
-- is the inverse matrix of couplings with $i,j=c,d,f$.
To calculate the internal
energy of SC state at $T_c$ we can use the mean-field Hamiltonian, which obeys $\av{H} \propto |\Delta|^2$. We diagonalize  the Hamiltonian (\ref{HBdG}), take average over the SC state, use the gap equation expression and obtain
\begin{equation}
\begin{gathered}
\av{H} =-\nu_c \int d\varepsilon_c \sqrt{\varepsilon_c^2 + \Delta_c^2} \tanh \frac{\sqrt{\varepsilon_c^2 + \Delta_c^2}}{2 T} - \nu_d \int d\varepsilon_d \sqrt{\varepsilon_d^2 + \Delta_d^2} \tanh \frac{\sqrt{\varepsilon_d^2 + \Delta_d^2}}{2 T} - \nu_c \int d\varepsilon_c \sqrt{\varepsilon_c^2 + \Delta_c^2} \tanh \frac{\sqrt{\varepsilon_c^2 + \Delta_c^2}}{2 T}+ \Delta_i \hat{g}^{-1}_{ij} \Delta_j = \\
=-\nu_c \int d\varepsilon_c \frac{\varepsilon_c^2 + \Delta_c^2 /2}{\sqrt{\varepsilon_c^2 + \Delta_c^2}} \tanh \frac{\sqrt{\varepsilon_c^2 + \Delta_c^2}}{2 T} - \nu_d \int d\varepsilon_d \frac{\varepsilon_d^2 + \Delta_d^2 /2}{\sqrt{\varepsilon_d^2 + \Delta_d^2}} \tanh \frac{\sqrt{\varepsilon_d^2 + \Delta_d^2}}{2 T} - \nu_f \int d\varepsilon_f \frac{\varepsilon_f^2 + \Delta_f^2 /2}{\sqrt{\varepsilon_f^2 + \Delta_f^2}} \tanh \frac{\sqrt{\varepsilon_f^2 + \Delta_f^2}}{2 T}.
\end{gathered}
\end{equation}
The solution (\ref{gap_eq_solsup}) of the gap equation allows to express $\Delta_d$ and $\Delta_f$ via $\Delta_c$:
\begin{equation}
\Delta_d = \alpha \Delta_c, \; \Delta_f= - \beta \frac{\nu_c}{\nu_f} \Delta_c,
\label{xynotsuppl}
\end{equation}
where $\alpha$ and  $\beta$ are defined in Eqs. (\ref{alpha_def}),(\ref{beta_def}). 
Close to $T_c$ we expand $\av{H}$ in powers of $\Delta_{c,d,f}$ and subtract the normal state energy to calculate the condensation energy in the notation of Eq. (\ref{xynotsuppl}):
\begin{equation}
\begin{gathered}
E_{cond} = \nu_c  |\Delta_c|^2 \int d\varepsilon_c \frac{\tanh^2 \frac{\varepsilon_c}{2 T_c} - 1}{4T_c} + \nu_d  |\Delta_d|^2 \int d\varepsilon_d \frac{\tanh^2 \frac{\varepsilon_d}{2 T_c} - 1}{4T_c}  + \nu_f  |\Delta_f|^2 \int d\varepsilon_c \frac{\tanh^2 \frac{\varepsilon_f}{2 T_c} - 1}{4T_c} = \\ = -\frac{\nu_c + \nu_d \alpha^2 + \nu_f (-\beta \frac{\nu_c}{\nu_f})^2 }{2} |\Delta_c|^2 =  - \Upsilon |\Delta_c|^2.
\end{gathered}
\end{equation}
Then the specific heat jump is given by
\begin{equation}
\frac{C_s - C_n}{C_n} = - \frac{3}{2\pi^2} \frac{\Upsilon}{(\nu_c + \nu_d + \nu_f) T_c} \frac{d |\Delta_c|^2}{d T},
\end{equation}
where close to $T_c$ the gap is given by the standard expression
$$
\Delta_c (T) \simeq 3.06 T_c \sqrt{\frac{\nu_c + \alpha^2 \nu_d +  (-\beta \frac{\nu_c}{\nu_f})^2 \nu_f}{\nu_c + \alpha^4 \nu_d +  (-\beta \frac{\nu_c}{\nu_f})^4 \nu_f}}  \sqrt{1- T/T_c},
$$
where the long fraction is obtained using Ginzburg-Landau expansion and shows the modification of gap magnitude compared to the conventional BCS value. 

\section{Special case when the specific heat jump is large}
\label{sec:param}

In this section we address the possibility of getting a large specific heat jump within the same model. As it can be seen from the calculations in Sec. \ref{sec:thermo}, one cannot reach $C_s-C_n \propto \nu_f$ in the system with regular magnitude of inter-pocket interactions $U_{fc}, U_{fd}$. Therefore, it
would be reasonable to look for large  $C_s-C_n$ in the system with \textbf{large} $U_{fc}, U_{fd} >> U_{cc},U_{dd},U_{cd}$. For simplicity we assume $U_{fc} = U_{fd}=U_{f1}$. Then instead of a 3-pocket model we can consider a 2-pocket model with indistinguishable inner $c,d$-fermion pockets, for which the $s$-wave gap equation matrix takes the form (here we absorbed the minus sign into the definition of interactions)
\begin{equation}
\begin{pmatrix}
\Delta \\
\Delta_f
\end{pmatrix}=L
\begin{pmatrix}
\nu U & \nu_f U_{f1}  \\
\nu U_{f1} & \nu_f U_{ff}
\end{pmatrix}
\begin{pmatrix}
\Delta \\
\Delta_f
\end{pmatrix},
\label{opp_gap_eq}
\end{equation}
where $U, \nu$ - is the interaction and DOS on two indistinguishable pockets respectively and $U_{f1}$ -- is the interaction between the inner pockets and the outer $d_{xy}$ pocket. This problem can be easily solved exactly:
\begin{equation}
\lambda \Delta = U \nu \Delta + U_{f1} \nu_f \Delta_f, \;\;\;  \lambda \Delta_f = U_{f1} \nu \Delta + U_{ff} \nu_f \Delta_f,
\label{opp_sol}
\end{equation}
where
\begin{equation}
\lambda = \frac{1}{2} \left( U \nu + U_{ff} \nu_f + \sqrt{(U \nu - U_{ff} \nu_f)^2 + 4 U_{f1}^2 \nu \nu_f} \right),
\end{equation}
is the largest eigenvalue of the gap equation matrix in (\ref{opp_gap_eq}). The specific heat jump depends on the gap ratio, which can be extracted from  Eq. (\ref{opp_sol}):
\begin{equation}
\left( \frac{\Delta}{\Delta_f} \right)^2 = \frac{\nu_f}{\nu} \frac{\lambda - U_{ff} \nu_f}{\lambda - U \nu}.
\end{equation}
Note, that the gap ratio is not strongly influenced by the value of $U_{f1}$. Superconducting state energy $E_{SC}$ for large
$\nu_f$ is then proportional to $\nu_f$:
\begin{equation}
E_{SC} \propto -\nu \frac{\nu_f}{\nu} \Delta_f^2 - \nu_f \Delta_f^2 \propto -\nu_f \Delta_f^2,
\end{equation}
which results in large value for the specific heat jump $(C_s-C_n)/C_n$.

Now one has to discuss the possible realization of this case. Typically, the inter-band interaction is weaker than the intra-band, so without enhancement $U_{f1}$ cannot be the dominant interaction. One of mechanisms of enhancement is the presence of spin fluctuations in the
system. Although spin fluctuations are usually expected to be damped in overdoped materials, they occur even in strongly hole-doped FeSCs like $\mathrm{K Fe_2 As_2}$ \cite{Lee2011PRL}.

In Ref. \cite{Lee2011PRL} authors reported observation of incommensurate spin fluctuations at finite momentum, which could be the ground for the inter-band interaction enhancement. However, the observed incommensurate momentum is far from the nesting vector within the hole pocket \cite{Lee2011PRL}. This allows for enhancement of interaction only at certain spots on the FS which are not "hot" in the cuprates language (the DOS at these spots is not increased). Thus the enhancement of inter-band interaction is not sufficient in bulk $\mathrm{K Fe_2 As_2}$, and experimental results don't show signatures of large specific heat jump.

Knowing that spin fluctuations arise from the magnetic order one can propose an experimental way of enhancing the inter-band interaction.
Consider a thin $\mathrm{K Fe_2 As_2}$ film \cite{Hiramatsu2014} on top of a magnetic substrate. The FS structure of $\mathrm{K Fe_2 As_2}$ film shouldn't differ from a bulk sample since this compound exhibit sufficiently 2D physics. Magnetic substrate should be prepared such, that the magnetic
vector $\bold{q}$ is equal to the nesting vector between the $d_{xz}/d_{yz}$ pockets and the $d_{xy}$ pocket. In the magnetic BZ, which contains 1 $\mathrm{Fe}$ atom, the flat $d_{xy}$ pocket is located not at $\Gamma$-point but at $(\pi, \pi)$. Hence, the nesting vector will be of the same order and $\bold{q}\simeq (\pi, \pi)$. Then the spin fluctuations originating from the bulk magnetism of the substrate will enhance the inter-band interactions in the sample.

\bibliography{biblio}